\documentclass[
reprint,
amsmath,amssymb,
aip,
jcp
]{revtex4-1}

\usepackage{natbib}
\usepackage{graphicx}
\usepackage{dcolumn}
\usepackage{bm}
\usepackage{hyperref}
\usepackage{color}
\usepackage{epstopdf}

\begin{document}
\title{Thermodynamics of two-dimensional Yukawa systems across coupling regimes}

\author{Nikita P. Kryuchkov}

\affiliation
{Bauman Moscow State Technical University,\\
2nd Baumanskaya str. 5, 105005 Moscow, Russia}

\author{Sergey A. Khrapak}
\affiliation
{Aix Marseille University, CNRS, PIIM, Marseille,  France; Institut f\"ur Materialphysik im Weltraum, Deutsches Zentrum f\"ur Luft- und Raumfahrt (DLR), Oberpfaffenhofen, Germany; Joint Institute for High Temperatures, Russian Academy of Sciences, Moscow, Russia}

\author{Stanislav O. Yurchenko}
\email{st.yurchenko@mail.ru}
\affiliation
{Bauman Moscow State Technical University,\\
2nd Baumanskaya str. 5, 105005 Moscow, Russia}

\date{\today}

\begin{abstract}
Thermodynamics of two-dimensional Yukawa (screened Coulomb or Debye-H\"uckel) systems is studied systematically using molecular dynamics (MD) simulations. Simulations cover very broad parameter range spanning from weakly coupled gaseous states to strongly coupled fluid and crystalline states. Important thermodynamic quantities such as internal energy and pressure are obtained and accurate physically motivated fits are proposed. This allows us to put forward simple practical expressions to describe thermodynamic properties of two-dimensional Yukawa systems. For crystals, in addition to numerical simulations, the recently developed shortest-graph interpolation method is applied to describe pair correlations and hence thermodynamic properties. It is shown that the finite-temperature effects can be accounted for by using simple correction of peaks in the pair correlation function. The corresponding correction coefficients are evaluated using MD simulation. The relevance of the obtained results in the context of colloidal systems, complex (dusty) plasmas, 
ions absorbed to interfaces in electrolytes is pointed out.
\begin{description}
\item[PACS numbers]
61.50.-f,
61.66.-f,
07.05.Tp
\end{description}
\end{abstract}

\maketitle
\section{Introduction}

Systems of particles interacting via Yukawa (screening Coulomb, or Debye-H\"uckel) potential are widely found in nature.
In the physics of soft matter, the Yukawa potential plays
particularly important role, because it is traditionally used to describe interactions between ions in screening media (for instance, in aqueous solutions of electrolytes), charged colloidal micro- and nano-particles in various solvents and at interfaces of fluid media,
as well as between charged particles in complex (dusty) plasmas.~\cite{C0SM00813C, ivlev.book, Loewen1994249, RevModPhys.81.1353}
In addition, Yukawa systems represent a useful model of classical interacting particles with the softness of interaction variable in a very wide range, from extremely soft Coulomb interaction (one-component-plasma limit~\cite{BausPR1980} in the absence of screening) to the extremely hard interaction (hard spheres~\cite{HS_Book} in the limit of very strong screening).

Various aspects of structural, dynamical, and thermodynamical properties of Yukawa systems were studied numerically, some representative examples can be found in Refs.~\onlinecite{1.453924,  1.459898, 1.464213, 1.472802, PhysRevE.56.4671, PhysRevE.72.026409, Vaulina20093330, qi:234508, 1.3524309,ToliasPRE2014}.
Results of these studies found wide applications, for instance, to explain phase transitions in complex (dusty) plasmas,~\cite{PhysRevLett.103.015001, PhysRevLett.105.115004,Klumov2010, PhysRevLett.106.205001,KhrapakPRE2012} and in colloidal systems.~\cite{Loewen1994249, 0953-8984-21-20-203101, PhysRevLett.58.1200, 1.2189850, la0340089, C2SM26473K}
In particular, in colloidal suspensions,
screened Coulomb repulsion determines various crystalline structures and their properties
(see, e.g., Refs.~\onlinecite{Loewen1994249, 0953-8984-21-20-203101, B704251P, C4SM02365J, Wang2015, PhysRevE.60.7157, PhysRevLett.74.4555}).
The screened Coulomb repulsion is the basic interaction for ions and microparticles in electrolytes. \cite{0034-4885-65-11-201, 0953-8984-12-46-201, C2SM26473K, B908331F, C3SM51752G, C2SM26729B}

Both three-dimensional (3D) and two-dimensional (2D) Yukawa systems can be of interest in the context of colloids, complex plasmas, and electrolytes. In the 2D situation the particles are normally  confined to a thin layer or are located at an interface.
For instance, 2D Yukawa systems of ions can arise in electrolytes at the interfaces due to the ion specific effects.
In drops of aqua solutions of electrolytes,
the ionic redistribution near the surface and surface trapping of anions change the surface tension
(see, e.g., Refs. \onlinecite{1.461592, jp012750g, PhysRevLett.103.257802, PhysRevLett.102.147803}).
In bulk aqua solutions of electrolytes,
similar ion-specific effects lead to the formation of  bubbles stabilized by ions, the so-called bubstons,~\cite{1.4739528, Bunkin2012, acs.jpcb.5b11103, acs.langmuir.6b01644}
in which the ions are adsorbed in thin layer inside the bubble surface to compensate the pressure by surface tension. Two-dimensional plasma crystals and fluids represent one of the major topics of experimental research into complex plasmas in laboratory conditions.~\cite{LinIPRL1994,ThomasPRL1994,Hayashi1994,MelzerPLA1994,
ThomasNat1996,MelzerPRE1996,Schweigert1998,Morfill2004,IvlevCPP2015}
Thus, 2D systems of Yukawa particles occur in a rather broad range of applications and related problems are of fundamental importance.

Thermodynamic properties of 2D Yukawa systems have been of considerable continuous interest in the last couple of decades, largely in the context of complex plasmas.~\cite{PhysRevE.70.016405, PhysRevE.72.026409, Vaulina20093330, VaulinaPLA_2014, KhrapakPoP08_2015, 0022-3727-49-23-235203, 1.4962685} However, to the best of our knowledge, no comprehensive results across coupling regimes along with simple and reliable approximations convenient for practical use have been proposed. In the present paper, we  study systematically  thermodynamics of 2D Yukawa systems in a very broad parameter regime, from very weakly interacting gaseous state to strongly interacting fluid and solid states. Using MD simulations we systematically calculate the excess energy and pressure. For gases and fluids simple practical expressions for the thermodynamic properties are then proposed. For crystals, the shortest-graph interpolation method~\cite{1.4869863, 1.4921223, 1.4926945, 0953-8984-28-23-235401} is applied to calculate the pair correlations and thermodynamic properties. The advantages of this method as well as its excellent accuracy are demonstrated.
Overall, this paper describes simple and reliable tools to calculate the thermodynamic properties of 2D Yukawa systems across coupling regimes with required accuracy.

\section{Methods}

\subsection{System description}
\label{SD}

We investigate a classical system of point-like particles in the 2D geometry interacting via the pairwise repulsive Yukawa potential of the form
\begin{equation*}
\varphi (r) = \frac{\varepsilon \lambda}{r}\exp\left(-\frac{r}{\lambda}\right),
\end{equation*}
where $\varepsilon$, and $\lambda$ are the energy and (screening) length scales of the interaction. For charged particles immersed in a plasma-like screening environment, the energy scale is $\varepsilon=Q^2/4\pi\epsilon_0\lambda$ (in SI units), where $Q$ is the charge and $\epsilon_0$ is the permittivity of free space. The properties of Yukawa systems are determined by the two dimensionless parameters. The first is the coupling parameter, $\Gamma = (Q^2/4\pi  \epsilon_0 a k_{\mathrm{B}}T)$, where $k_{\mathrm{B}}$ is the Boltzmann constant, $T$ is the temperature, $a=(\pi  n)^{-1/2}$ is the 2D Wigner-Seitz radius, and $n=N/V$ is the areal density of $N$ particles occupying the 2D volume $V$. The second is the screening parameter, $\kappa = a/\lambda$. Note, that the coupling parameter is roughly the ratio of the potential energy of interaction between two neighbouring particles to their kinetic energy. The system is usually said to be in the strongly coupled state when this ratio is large, that is $\Gamma\gtrsim 1$.

When coupling increases the system forms a strongly coupled fluid phase, which can crystallize upon further increase in $\Gamma$. This fluid-solid transition can be characterized by the temperature and/or coupling parameter,  $T_{\rm m}$ and $\Gamma_{\rm m}$, where the subscript ``m'' refers to melting. Both $T_{\rm m}$ and $\Gamma_{\rm m}$ are the functions of the screening parameter $\kappa$. The dependence $\Gamma_{\rm m}(\kappa)$ has been approximated in Ref.~\onlinecite{PhysRevE.72.026409} by the following fit:
\begin{equation}\label{Melting2D}
\Gamma_{\rm m}(\kappa)\simeq \frac{131}{1-0.388\kappa^2+0.138\kappa^3-0.0138\kappa^4}.
\end{equation}
This fit describes relatively well the melting points found from the bond angular correlation analysis (see Fig.~6 of Ref.~\onlinecite{PhysRevE.72.026409}) up to $\kappa = 3.0$ and it should  not be applied for larger $\kappa$. In the limit $\kappa = 0$ the system reduces to the 2D one-component-plasma (OCP) with the Coulomb interaction. In this case $\Gamma_{\rm m}\simeq 131$ lies in the range predicted in earlier numerical simulations~\cite{Gann1979} and obtained in experiments with a classical 2D sheet of electrons~\cite{Grimes1979} (see also Ref.~\onlinecite{KhrapakCPP2016} for a recent overview of OCP thermodynamics in 2D and 3D).

Finally, it is worth to comment on the nature of the fluid-solid phase transition in 2D Yukawa systems. Recently, it has been demonstrated that the potential softness is very important factor, which determines the melting scenario.~\cite{KapferPRL2015}
For sufficiently steep repulsive interactions the hard-disk melting scenario holds: a first-order liquid-hexatic and a continuous
hexatic-solid transition can be identified. ~\cite{PhysRevLett.107.155704, PhysRevE.87.042134} For softer interactions the liquid-hexatic transition is continuous, with correlations consistent with the Kosterlitz-Thouless-Halperin-Nelson-Young (KTHNY) scenario.  (For example, in 2D colloidal systems, hexatic phase was observed in the experiment by Zahn et al.~\cite{PhysRevLett.82.2721}) For the Yukawa potential the transition between these two scenarios occurs at about $\kappa\simeq 6$.~\cite{KapferPRL2015} Below we consider systems with $\kappa$ in the range from $0.5$ to $3.0$ (this range is particularly relevant to 2D plasma crystals and fluids in laboratory experiments~\cite{FortovUFN2004,FortovPR2005},\cite{CTPP:CTPP201400099}), thus belonging to the soft interaction class.  In this range of $\kappa$, the hexatic phase occupies a rather narrow region on the phase diagram,~\cite{KapferPRL2015} and the study of its properties is beyond the scope of the present investigation.

\subsection{Computational details}
\label{MDdetails}

To obtain the thermodynamic properties of the 2D Yukawa systems across coupling regime, extensive MD simulations have been performed. The MD simulations have been done in the $NVT$ ensemble at different temperatures using $N=64 000$ particles and the Langevin thermostat. The numerical time step was chosen $\Delta t_c=5\times 10^{-4}\sqrt{m\lambda^2/\epsilon}$ for the crystalline phase and $\Delta t_c \sqrt{\Gamma/\Gamma_{\rm m}}$  for the fluid phase. The cutoff radius of the Yukawa potential was set equal to $15n^{-1/2}$. The simulations were run for $1.5\times 10^6$ time steps to equilibrate the system and obtain the equilibrium properties. In the simulation run with $\kappa = 0.5$ Ewald summation was implemented.

The simulations have been performed for a number of screening parameters $\kappa$ ranging from $0.5$ to $3.0$. This corresponds to sufficiently soft interactions as discussed above. For each value of the screening parameter $\kappa$, twelve simulation runs correspond to the fluid phase and nine runs to the crystalline phase. In the fluid phase the coupling parameter ranges from $\Gamma=0.5$ to $\simeq 0.95\Gamma_{\rm m}$. In the solid phase the values corresponding to $\Gamma_{\rm m}/\Gamma=0.9,0.8,...,0.1$ are taken.

The main simulation results are summarized in Tables~\ref{Table1}-\ref{Table4} of the Appendix.


\subsection{Thermodynamic definitions and relations}\label{Thermo}

The main thermodynamic quantities which will be required below are the internal energy $U$, Helmholtz free energy $F$, and pressure $P$ of the system. The following thermodynamic definitions exist~\cite{LL}
\begin{eqnarray}
U=-T^2\left(\frac{\partial}{\partial T}\frac{F}{T}\right)_V, \\
P=-\left(\frac{\partial F}{\partial V}\right)_T.
\end{eqnarray}
In addition, $U$ and $P$ can be calculated using the integral equations of state~\cite{hansen-book, frenkel2001}
\begin{equation}
\begin{split}
& U= N\left(k_{\rm B}T+ n\int{d\mathbf{r}\; \varphi(r)g(\mathbf{r})}\right),\\
& PV = N\left(k_{\rm B}T - \frac{n}{4}\int{d\mathbf{r}\; r\varphi'(r)g(\mathbf{r})} \right),
\end{split}
\end{equation}
where $g(\mathbf{r})$ denotes the radial distribution function, which is isotropic in gas and fluid phases and anisotropic in the crystalline phase.

We will use conventional reduced units: $u=U/Nk_{\rm B}T$, $f=F/Nk_{\rm B}T$, and $p=PV/Nk_{\rm B}T$ and divide the thermodynamic quantities into the kinetic (ideal gas) and potential (excess) components, so that $u=1 + u_{\rm ex}$ (in 2D), $f=f_{\rm id}+f_{\rm ex}$, and $p=1+p_{\rm ex}$. Finally, it is useful to operate with the Yukawa system state variables $\Gamma$ and $\kappa$. In these variables the thermodynamic identities for 2D Yukawa fluids are~\cite{KhrapakPoP08_2015, 1.4935846}
\begin{equation}\label{pf}
p=1+\frac{\Gamma}{2}\frac{\partial f_{\rm ex}}{\partial\Gamma}-\frac{\kappa}{2}\frac{\partial f_{\rm ex}}{\partial\kappa}, \qquad
f_{\rm ex} = \int_0^{\Gamma}{d\Gamma'\; \frac{u_{\mathrm{ex}}(\kappa, \Gamma')}{\Gamma'}}.
\end{equation}

\subsection{The shortest-graph method}

To describe the thermodynamics of 2D Yukawa crystals analytically,
we employ the shortest-graph method, proposed and developed in Refs.~\onlinecite{1.4869863, 1.4926945, 0953-8984-28-23-235401}.
Following these papers, thermodynamical properties of classical crystals can be obtained very accurately from the following consideration. The anisotropic pair-correlation function $g(\mathbf{r})$ of a crystal is written in the form
\begin{equation}
\label{Eq1}
g(\mathbf{r}) = \frac{1}{n}\sum_\alpha{p_\alpha(\mathbf{r}-\mathbf{r_\alpha})},
\end{equation}
where the summation is over all the nodes $\alpha$, and
each individual peak has the shape
\begin{equation}
\label{Eq2}
\begin{split}
&p_\alpha(\mathbf{r}) \propto
 \exp\left[-\frac{\varphi(\mathbf{r}+\mathbf{r_\alpha})}{k_{\rm B}T}-b_\alpha(\mathbf{e_\alpha}\cdot\mathbf{r})-
\right. \\
& \qquad\qquad \qquad \qquad \left.
-\frac{(\mathbf{e_\alpha}\cdot\mathbf{r})^2}{2 a_{\|\alpha}^2}-
\frac{\mathbf{r}^2-(\mathbf{e_\alpha}\cdot\mathbf{r})^2}{2 a_{\perp\alpha}^2}\right].
\end{split}
\end{equation}
The normalization constant as well as the parameters $a_{\|,\perp\alpha}^2, b_\alpha$ are defined by the following conditions\cite{0953-8984-28-23-235401}
\begin{equation}
\label{Eq3}
\begin{split}
& \int{d\mathbf{r}\;p_\alpha(\mathbf{r})}=1, \qquad \int{d\mathbf{r}\;\mathbf{r}p_\alpha(\mathbf{r})}=0, \\
& \int{d\mathbf{r}\;(\mathbf{e_\alpha}\cdot\mathbf{r})^2 p_\alpha(\mathbf{r})}=\sigma_{\|\alpha}^2,\\
& \int{d\mathbf{r}\;[\mathbf{r}^2-(\mathbf{e_\alpha}\cdot\mathbf{r})^2] p_\alpha(\mathbf{r})}=(D-1) \sigma_{\perp\alpha}^2,
\end{split}
\end{equation}
where $D=2$ is the spatial dimensionality and $\mathbf{e_\alpha}=\mathbf{r_\alpha}/r_\alpha$ is the unit vector in the
direction of $\mathbf{r_\alpha}$,
$\sigma_{\|,\perp}^2$ is the mean squared displacement for longitudinal and transversal directions, respectively, calculated using the finite-temperature phonon spectra,
taking into account the anharmonic effects.\cite{0953-8984-28-23-235401} By using the pair correlation function $g(\mathbf{r})$ the excess energy and pressure can then be obtained.
However, calculation of the finite-temperature phonon spectra is a difficult problem, which is beyond the scope of the present paper.
Therefore, we propose here a simpler practical approach, which yields very accurate results and can be used for practical calculations.

Due to the anharmonicity of phonon spectra at finite temperatures,
the second-order term becomes more significant in the temperature expansion of the mean-squared displacements $\sigma^2$.
To account for this effect, we propose the anharmonic correction of the mean-squared displacements
\begin{equation}
\label{Eq5}
\sigma_{\|,\perp\alpha}^2 = \widetilde{\sigma}_{\|,\perp\alpha}^2 \left[1+\beta(\kappa)N\widetilde{\sigma}_{1}^2/V\right],
\end{equation}
where the tildes denote the mean-squared displacement calculated using zero-temperature phonon spectra (see Ref.\onlinecite{1.4926945}),
$\widetilde{\sigma}_1^2$ is the total mean-squared displacement for the nearest neighbours, and we have introduced the anharmonic correction coefficient $\beta(\kappa)$, which does not depend on the temperature and should be found using MD simulations for different screening parameters.
The correction given by Eq.\eqref{Eq5} conserves the ratio $\sigma_\|^2/\sigma_\perp^2$ between the mean-squared displacements in the longitudinal and transversal directions.
\emph{A posteriori} comparison with MD results proves that this assumption allows to obtain excellent accuracy.

\section{Results}

\subsection{Weakly-coupled fluids}

\begin{figure}[!b]
    \centering
    \includegraphics[width=85mm]{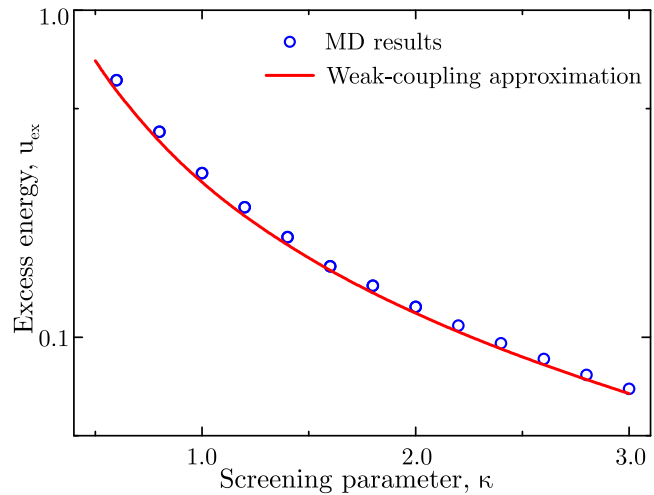}\\
    \caption{The excess energy $u_{\rm ex}$ of 2D Yukawa weakly coupled fluids versus the screening parameter $\kappa$ at a fixed coupling parameter $\Gamma = 0.5$. The symbols correspond to the results of MD simulations, the solid curve is plotted using the analytical expression of Eq.~(\ref{SVC}).
    }
\label{FigSC}
\end{figure}

A simple and physically transparent approach to the thermodynamics of weakly coupled Yukawa systems for small deviations from the ideal gas behavior is to calculate the  second virial coefficient. This has recently been shown to work well in 3D Yukawa systems.~\cite{KhrapakPPCF2016} In the 2D geometry the excess free energy is expressed in this approximation as
\begin{equation}\label{SVC}
f_{\rm ex}\simeq \pi n \int\left[1-e^{-\varphi(r)/k_{\rm B}T}\right]r dr.
\end{equation}
The excess energy and pressure can be readily obtained from the excess free energy. We compare the values $u_{\rm ex}$ at a fixed coupling parameter $\Gamma=0.5$ obtained from Eq.~(\ref{SVC}) and computed using MD simulations in Fig.~\ref{FigSC}. The agreement is satisfactory: in the range of $\kappa$ investigated the deviations are within several percent. The agreement naturally improves with increasing $\kappa$, because at a fixed $\Gamma$ the actual interaction strength weakens as $\kappa$ increases.

\subsection{Strongly-coupled fluids}

The excess energy and pressure of the 2D Yukawa fluids have been determined using MD simulations in a wide range of coupling and screening parameters. The results are summarized in the Table \ref{Table1} of the Appendix. Here we describe simple analytical
approximations, which can be used to evaluate the energy and pressure for practical purposes.

In the strongly coupled fluid regime it is helpful to divide the thermodynamic quantities, such as energy and pressure, into static and thermal contributions. The static contribution corresponds to the value of internal energy when the particles are frozen in some regular configuration and the thermal corrections arise due to the deviations of the particles from these fixed positions (due to thermal motion). Of course, such a division is only meaningful when the regular structure is specified. For crystals, the obvious choice is a corresponding lattice sum (Madelung energy). For fluids this choice is also meaningful and we use it here (Note, that in 3D Yukawa system a slightly different definition of the static fluid energy is traditionally employed.~\cite{KhrapakPPCF2016, KhrapakISM})

\begin{table}[!b]
\caption{\label{TabM} Madelung constants of the 2D Yukawa crystals (triangular lattice) for various screening parameters in the range $0.5 \leq \kappa\leq 3.0$ }
\begin{ruledtabular}
\begin{tabular}{cccc}
$\kappa$ & $M$ & $\kappa$ & $M$   \\ \hline
0.5 & 1.11914 & 1.8 & 0.05449  \\
0.6 & 0.82503 & 2.0 & 0.03660  \\
0.8 & 0.48127 & 2.2 & 0.02470  \\
1.0 & 0.29709 & 2.4 & 0.01672  \\
1.2 & 0.18960 & 2.6 & 0.01135  \\
1.4 & 0.12357 & 2.8 & 0.00772  \\
1.6 & 0.08167 & 3.0 & 0.00525 \\ 
\end{tabular}
\end{ruledtabular}
\end{table}

The excess internal energy is thus a sum of the static and thermal contributions,
\begin{equation}
u_{\rm ex} = u_{\rm st} + u_{\rm th},
\end{equation}
where $u_{\rm st} = M\Gamma$ and $M$ is the Madelung constant.
The values of the Madelung constant for 2D Yukawa systems in the regime of relatively weak screening, $0.5 \leq \kappa\leq 3.0$, are tabulated in Table~\ref{TabM}. The dependence $M(\kappa)$ can be fitted using a functional form similar to that proposed by  Totsuji~\emph{et al.}\cite{PhysRevE.70.016405}
\begin{equation}
\label{Eq6}
M = -1.1061+0.5038\kappa-0.11053\kappa^2+0.00968\kappa^3+1/\kappa.
\end{equation}
The last term in (\ref{Eq6}) accounts for the absence of neutralizing background in our case (but present in Ref.~\onlinecite{PhysRevE.70.016405}), the energy of this background being simply $-\Gamma/\kappa$. The fit is chosen in such a way that when $\kappa\rightarrow 0$ and the neutralizing background is introduced, the Madelung constant is reduced to the well known value of the triangular lattice sum of the 2D one-component-plasma (OCP) with Coulomb interactions, $M_{\rm OCP}\simeq -1.1061$. This fit is accurate to within a tiny fraction of percent for $\kappa\lesssim 1.0$ and to within $\sim 1\%$ when screening becomes stronger ($\kappa\sim 3$).

\begin{figure}[!t]
    \centering
    \includegraphics[width=85mm]{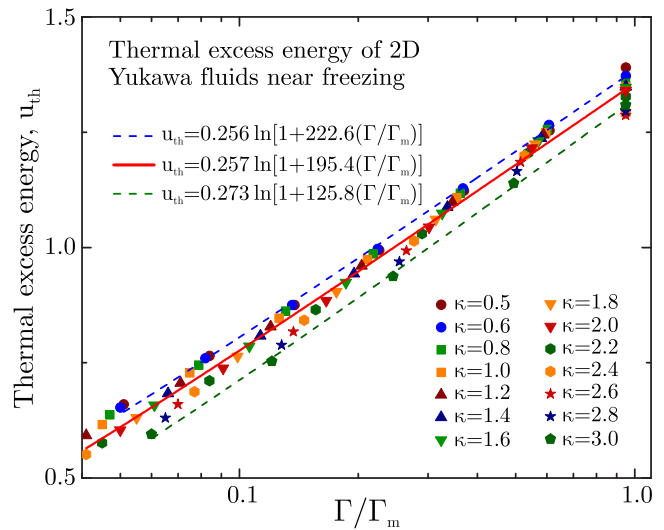}\\
    \caption{
    Thermal component of the reduced excess energy, $u_{\rm th}$ of 2D Yukawa fluids near the fluid-solid phase transition versus the reduced coupling parameter $\Gamma/\Gamma_{\rm m}$. Symbols correspond to MD simulations for different values of the screening parameter $\kappa$. The curves are the analytical fits to these data using Eq.~(\ref{Fit1}): The upper (lower) curve corresponds to fitting the MD results for $\kappa=0.5$ ($\kappa = 3.0$) and the intermediate (red) curve is obtained by fitting the entire massive of the data points.}
\label{FigR1}
\end{figure}

The thermal part of the excess energy is expected to exhibit a quasi-universal scaling with respect to the reduced coupling parameter $\Gamma/\Gamma_{\rm m}$. This is a general property of classical particle systems with sufficiently soft interactions, which was first pointed out by Rosenfeld and Tarazona (RT scaling) for 3D systems.~\cite{RT1,RT2} In the context of 3D Yukawa systems, the RT scaling has been proven to be very useful in Refs.~\onlinecite{1.4921223,KhrapakPPCF2016,KhrapakPRE2015,KhrapakPRE03_2015} The emergence of RT scaling analogue for 2D systems has been discussed in the context of OCP with Coulomb and logarithmic interactions, Yukawa systems near the OCP limit, and inverse-power-law interactions.~\cite{KhrapakCPP2016,KhrapakPoP08_2015} The dependence of $u_{\rm th}$ on $\Gamma/\Gamma_{\rm m}$ in the strongly coupled regime is displayed in Fig.~\ref{FigR1}. The quasi-universality is well pronounced, although there is clearly some systematic tendency of decreasing the value of $u_{\rm th}$ with $\kappa$ at the same value of  $\Gamma/\Gamma_{\rm m}$. This tendency is expected when the potential steepness increases (see e.g. Fig.~4 from Ref.~\onlinecite{KhrapakPoP08_2015}). Overall, the data points corresponding to the dependence $u_{\rm th}(\Gamma/\Gamma_{\rm m})$ are confined to a relatively narrow range. The important point is that towards the side of soft interactions (sufficiently small $\kappa$ in our case), the static component of the internal energy is dominant over the thermal one. For example, at $\kappa=1$ the thermal component contributes only to about $2\%$ of the total excess energy near the fluid-solid phase transition. Therefore, even moderately accurate fits for $u_{\rm th}$ allow to obtain high accuracy with respect to the total excess energy $u_{\rm ex}$.

Three fits are shown in Fig.~\ref{FigR1}. The upper (lower) curve corresponds to the data portion for $\kappa=0.5$ ($\kappa = 3.0$).
The intermediate curve has been obtained using the entire massive of the data points (corresponding to the parameter regime shown). It can be considered as representative for strongly coupled 2D Yukawa fluids in the vicinity of the freezing transition.
The functional form of the fit is the same as used previously~\cite{KhrapakPoP08_2015}
\begin{equation} \label{Fit1}
u_{\rm th} =A \ln (1+B\Gamma/\Gamma_{\rm m}).
\end{equation}
The use of the coefficients $A=0.257$ and $B=195.4$ determined here would somewhat improve previous approximations.

The excess free energy can be routinely calculated using the model for the excess energy formulated above and the second of Eqs.~(\ref{pf}). The resulting expression is rather simple,
\begin{equation}\label{fex}
f_{\rm ex}=M(\kappa)\Gamma - A{\rm Li}_2(-B\Gamma/\Gamma_{\rm m}),
\end{equation}
where ${\rm Li}_2(z)=\int_z^0 dt \ln(1-t)/t$ is dilogarithm. Note that in deriving Eq.~(\ref{fex}), the thermodynamic integration over the coupling parameter from 0 to $\Gamma$ has been performed, while Eq.~(\ref{Fit1}) is strictly speaking not applicable at $\Gamma\ll 1$.
The correct procedure would be to start thermodynamic integration from some small, but finite value $\Gamma_0$, and then add the constant $f_{\rm ex}(\Gamma_0)$ evaluated using Eq.~(\ref{SVC}). However, since the actual contribution from the weakly coupling regime is small, Eq.~(\ref{fex}) remains rather accurate at strong coupling and we use it here.

The calculation of pressure from the excess free energy is straightforward, but rather cumbersome in the considered case. This is because the differentiation with respect to $\kappa$ is involved, and  the two fits for $M(\kappa)$ and $\Gamma_{\rm m}(\kappa)$ are present. For this reason, the explicit expression for $p$ is not displayed. We verified that near freezing (at $\Gamma/\Gamma_{\rm m}\simeq 0.95$) the derived expression yields the pressures which deviate from the exact MD results by $\sim 0.001\%$ at $\kappa=0.5$, $~\sim 0.1\%$ at $\kappa=1.0$, and $\sim 1\%$ at $\kappa = 2.0-2.8$. The accuracy drops at the highest value $\kappa=3.0$. This is not surprising, since the fits for $M(\kappa)$ and $\Gamma_{\rm m}(\kappa)$ are only applicable for $\kappa\lesssim 3.0$ and, therefore, derivatives from these fits at   $\kappa=3.0$ can produce significant errors.

\begin{figure}[!t]
    \centering
    \includegraphics[width=85mm]{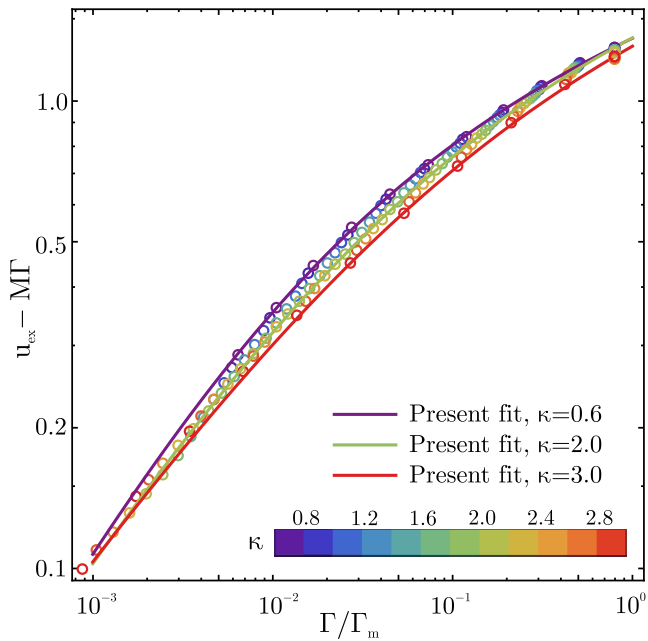}\\
    \caption{Dependence of the excess thermal energy $u_{\rm th}$ on the reduced coupling parameter $\Gamma/\Gamma_{\rm m}$. All the data points from numerical simulations are plotted. Solid curves correspond to three representative fits using Eq.~\eqref{Eq7}.}
\label{FigR2}
\end{figure}

We also found out that if better accuracy is required, the data for the excess thermal energy can be fitted by the following slightly modified expression
\begin{equation}
\label{Eq7}
u_{\mathrm{ex}} = A(\kappa)\ln\left[ 1 + B(\kappa) \Gamma^{s(\kappa)} \right],
\end{equation}
where $A$ and $B$ are now assumed $\kappa$-dependent and a $\kappa$-dependent exponent $s$ is introduced. Based on all the data points obtained in MD simulations the following relations are identified:
$A(\kappa) = 0.35708 + 0.09397\kappa$,
$B(\kappa)= 1.65491\exp(- 0.76911\kappa)$,
$s(\kappa) = 0.68838 - 0.05183\kappa$.
Some representative examples are shown in Fig.~\ref{FigR2}.
The fit of Eq.~(\ref{Eq7}) is clearly more accurate and can be used in
the regime of weaker coupling, compared to the simple form (\ref{Fit1}). However, it is also less practical in evaluating thermodynamic parameters other than the excess internal energy.

\subsection{Relation between excess pressure and energy}

It is sometimes advantageous to operate with an equation of state written in the form of relation between the pressure and internal energy of the system. For soft purely repulsive potentials a simplest formulation of this kind can be written as
\begin{equation}\label{gamma_ex}
p_{\rm ex}=\gamma_{\rm ex}u_{\rm ex}.
\end{equation}
Here the parameter $\gamma_{\rm ex}$ generally depends both on the temperature and density, that is both on $\Gamma$ and $\kappa$ for Yukawa systems. Note that the parameter $\gamma_{\rm ex}$ introduced in this way is not directly related to the conventional definitions of either the density scaling exponent or Gr\"uneisen parameter.~\cite{HummelPRB2015} Nevertheless, it may be helpful in characterizing the softness of the repulsive potential. We remind that for inverse-power-law (IPL) repulsive potentials of the form $\varphi(r)\propto r^{-\alpha}$ the relation between the excess pressure and energy is particularly simple, $p_{\rm ex}=\tfrac{\alpha}{2} u_{\rm ex}$ in 2D. Thus, an ``effective IPL exponent'' may be associated with the quantity $2\gamma_{\rm ex}$.

\begin{figure}[!t]
    \centering
    \includegraphics[width=85mm]{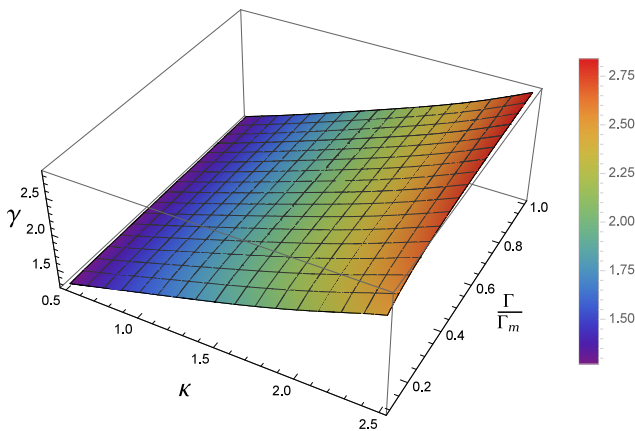}\\
    \caption{Ratio of the excess pressure to the excess energy, $\gamma_{\rm ex}=p_{\rm ex}/u_{\rm ex}$ on the plane ($\kappa$, $\Gamma/\Gamma_{\rm m}$).
    }
\label{gamma}
\end{figure}

Having approximations for both $p_{\rm ex}$ and $u_{\rm ex}$ for 2D Yukawa fluids we can easily estimate the value of $\gamma_{\rm ex}$. The corresponding plot of $\gamma_{\rm ex}$ as a function of Yukawa systems state variables $\kappa$ and $\Gamma/\Gamma_{\rm m}$ is shown in Fig.~\ref{gamma}. To produce this plot, Eq.~(\ref{Fit1}) for the thermal component of the excess energy has been used. Figure~\ref{gamma} shows that in the strongly coupled regime $\gamma_{\rm ex}$ is very weakly dependent on the coupling strength (temperature), but exhibits considerable dependence on $\kappa$ (density). Using the exact MD results for $p_{\rm ex}/u_{\rm ex}$ in the vicinity of the fluid-solid phase transition ($\Gamma/\Gamma_{\rm m}\simeq 0.95$) we have obtained a representative dependence $\gamma_{\rm ex}(\kappa)$ in the strongly coupled regime:
\begin{equation}
\gamma_{\rm ex}(\kappa)=1+0.526\kappa+0.13\kappa^2-0.02\kappa^3.
\end{equation}
Importantly, $\gamma_{\rm ex}\rightarrow 1$ as $\kappa\rightarrow 0$.
This seems counter-intuitive at first, because one would naturally expect $\gamma_{\rm ex}=\tfrac{1}{2}$ in the OCP Coulomb interaction limit in 2D. The difference is attributed to the presence of the neutralizing background in the OCP model. In the limit of very soft interaction, the energy and pressure are dominated by their static contributions. As $\kappa\rightarrow 0$, the dominant contribution is the Madelung energy, so that $f_{\rm ex}\sim u_{\rm ex}\sim M\Gamma\sim \Gamma/\kappa$ (without background). This implies $p_{\rm ex}=\tfrac{\Gamma}{2}(\partial f_{\rm ex}/\partial \Gamma)-\tfrac{\kappa}{2}(\partial f_{\rm ex}/\partial \kappa)\sim \Gamma/\kappa\sim u_{\rm ex}$. In the presence of neutralizing background the term $\Gamma/\kappa$ disappears and we have $f_{\rm ex}\sim u_{\rm ex}\sim M_{\rm OCP}\Gamma$. This yields $p_{\rm ex}\sim \tfrac{1}{2}M_{\rm OCP}\Gamma\sim \tfrac{1}{2}u_{\rm ex}$. This consideration demonstrates that the Yukawa systems in the limit $\kappa\rightarrow 0$ are not fully equivalent to the Coulomb systems with the neutralizing background.

\subsection{Crystals}

In a series of MD simulations for 2D Yukawa crystals, in addition to evaluate the excess energy and pressure (which are summarized in Tables~\ref{Table3} and \ref{Table4} of the Appendix), the mean squared displacements were calculated to find the anharmonic correction coefficient $\beta$. The resulting dependence $\beta(\kappa)$ is shown in Figure~\ref{FigR3} (the corresponding values are also tabulated in Table~\ref{Table2} of the Appendix for completeness).
The inset in Fig.~\ref{FigR3} presents the radial (isotropic) pair correlation function, $g(r) \propto \int{d\varphi\; g(\mathbf{r})}$,
and demonstrates excellent representation of the short- and long-distance correlations. The obtained anharmonic correction coefficient $\beta(\kappa)$ allows to calculate analytically pair correlation function and then the excess energy, pressure and other thermodynamic parameters by the thermodynamic integration with the help of the expressions given in Sec.~\ref{Thermo}.

\begin{figure}[!t]
    \centering
    \includegraphics[width=85mm]{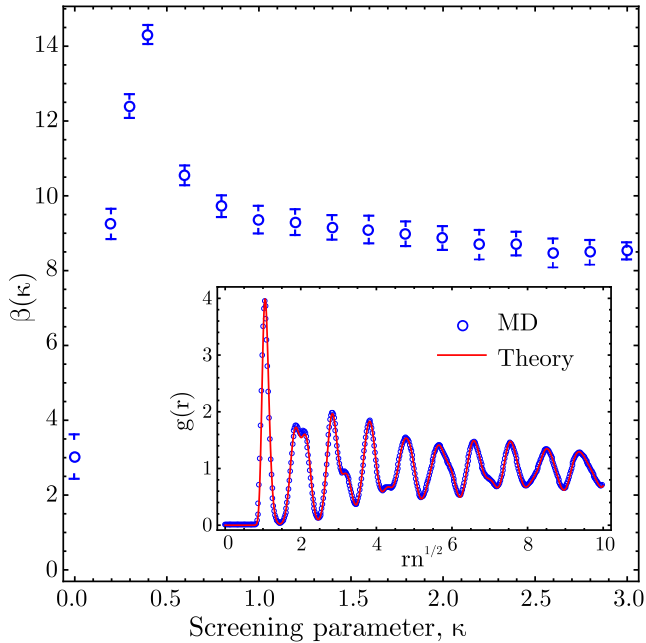}\\
    \caption{    Dependence of the anharmonic correction coefficient $\beta$ on the screening parameter $\kappa$. The inset demonstrates a typical comparison between the radial distribution functions obtained in a direct MD simulation and computed using the shortest-graph method. For details see the text.}
\label{FigR3}
\end{figure}

It is worth to point out the following observation:
In the limit $\kappa \rightarrow 0$, the Yukawa interaction tends to the unscreened Coulomb interaction $\varphi \propto r^{-1}$. According to our previous MD simulations,~\cite{1.4926945}
the finite-temperature phononic spectra differ weakly from zero-temperature ones for IPL potentials, $\varphi \propto r^{-\alpha}$. Therefore, in the OCP limit ($\kappa=0$ and $\alpha=1$) we should obtain the smallest values of $\beta(\kappa)$. This is indeed observed in Fig.~\ref{FigR3}.

\begin{figure}[!t]
    \centering
    \includegraphics[width=85mm]{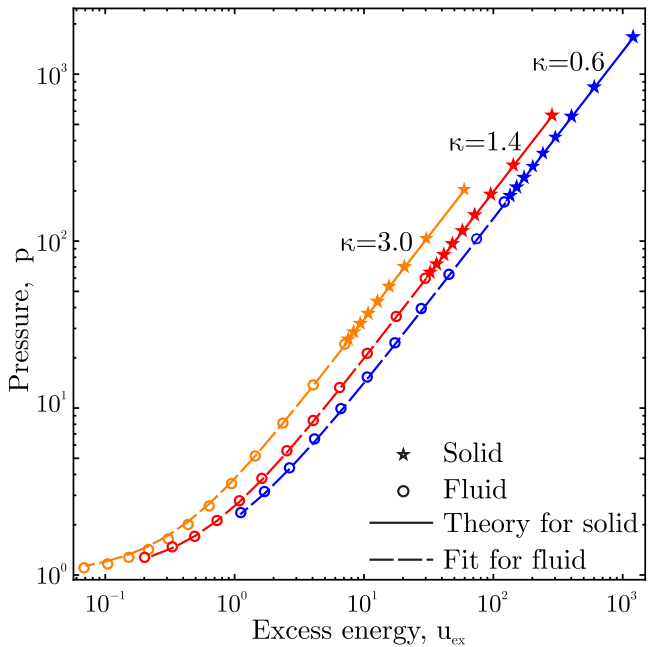}\\
    \caption{ Dependence of the reduced pressure on the reduced excess energy. Open (solid) symbols are the results of MD simulations for fluids and solids, respectively. The solid and dashed curves correspond to the shortest-graph method for solids and to the fit of Eq.~(\ref{Eq7}) for fluids.}
\label{FigR4}
\end{figure}

In figure~\ref{FigR4} we plot the reduced pressure versus the reduced excess energy of 2D Yukawa fluids and solids. Symbols are the MD results, the solid and dashed curves correspond to the shortest-graph method [with found anharmonic correction coefficient $\beta(\kappa)$] for the crystalline phase and the proposed fit by Eq.\eqref{Eq7} for the fluid phase, respectively. Excellent agreement is observed.

\subsection{Accuracy}

The relative difference between the excess energies calculated using the shortest-graph method and those evaluated using direct MD simulations in the solid phase amounts to $\simeq5\times 10^{-5}$, which is comparable to the values reported earlier.~\cite{0953-8984-28-23-235401} The accurate fit of Eq.~\eqref{Eq7}
yields the relative error in the excess energy smaller than $5\times10^{-4}$ and  $2\times10^{-3}$  for 72\% and 95\% of
the examined fluids data points, respectively. Maximal relative deviation, $5\times 10^{-3}$, is observed near the melting line at large values of the screening parameter $\kappa$. A simpler fit of Eq.~(\ref{Fit1}) is applicable when the relative deviations within $\lesssim 1\%$ are acceptable.

\begin{figure}[!t]
    \centering
    \includegraphics[width=85mm]{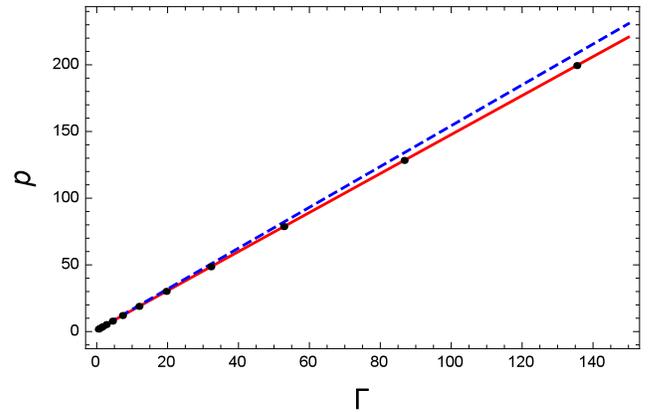}\\
    \caption{Reduced pressure, $p$, as a function of the coupling parameter $\Gamma$ for a Yukawa 2D fluid with the screening parameter $\kappa=0.5$. The symbols are exact MD results, the solid (red) line corresponds to the fit of Eq.~(\ref{Fit1}), the dashed (blue) line is the fit from Ref.~\onlinecite{0022-3727-49-23-235203}.}
\label{FigPressure}
\end{figure}

In addition, we can compare our results with those recently reported in Refs.~\onlinecite{0022-3727-49-23-235203,1.4962685}, where fits for the pressure of 2D Yukawa fluids in the $(\kappa,\Gamma)$ parameter space have been proposed. The case $\kappa=0.5$ received special attention and a simple two-term fit has been proposed based on the results of a MD simulation,~\cite{0022-3727-49-23-235203} $p=1.53\Gamma+1.33$.
We plot our MD results along with the fit of Eq.~(\ref{Fit1}) and the fit from Ref.~\onlinecite{0022-3727-49-23-235203} in Fig.~\ref{FigPressure}. One can see that the fit from Ref.~\onlinecite{0022-3727-49-23-235203} overestimates the pressure systematically at high values of $\Gamma$. At the strongest coupling in the fluid phase studied in this work, $\Gamma=135.42$, the present MD simulation yields $p= 199.434$, fit by Eq.~(\ref{Fit1}) yields $p=199.432$, while the fit from Ref.~\onlinecite{0022-3727-49-23-235203} yields $p=208.523$. On the other hand, the previous model for 2D Yukawa systems in the OCP (weakly screening) limit discussed in Refs.~\onlinecite{KhrapakPoP08_2015,1.4935846}
yields $p=199.445$, providing confidence in the accuracy of the  present results. The reasons for deviations in Ref.~\onlinecite{0022-3727-49-23-235203} have to be identified.

\section{Conclusion}

We studied thermodynamics of 2D classical Yukawa systems across coupling regimes, from the weakly non-ideal gas to the strongly coupled fluid and crystalline phases. Careful analysis of the extensive MD simulation results allowed us to propose simple and physically suitable expressions for the internal energy and pressure of the studied systems. These expressions can be used to estimate other thermodynamic properties using the conventional thermodynamic relations.

For weakly non-ideal gases, virial expansion with the two first terms retained was shown to provide reasonably good estimate of the excess  energy at sufficiently weak coupling, $\Gamma\lesssim 1$.

For the strongly-coupled fluid phase, we made use of the 2D analogue of the 3D Rosenfeld-Tarazona quasi-universal scaling of the thermal component of the excess energy. This quasi-universal scaling was shown to be particularly useful on approaching the fluid-solid phase transition. Deviations from the quasi-universal behaviour have been discussed and quantified.

To calculate thermodynamic properties of the crystalline phase, we employed the shortest-graph method for pair correlation functions. To account for the effects of finite-temperature phonon spectra, we proposed a simple way to correct the mean squared displacements of correlation peaks for different nodes. 
The coefficient of anharmonic correction was evaluated in MD simulations and then used in analytical estimates. The efficiency and accuracy of the approach was documented.


The results of this paper can be useful for thermodynamic calculations related to various phenomena in 2D and quasi-2D Yukawa fluids and solids in a broad range of parameters. In particular, this includes colloidal systems, complex (dusty) plasmas, and aqua solutions of electrolytes for ionic fluids.

\acknowledgments
The numerical simulations are supported by Russian Science Foundation, Project No. 14-29-00277. 
Post-processing is supported by Russian Foundation for Basic Researches (Projects Nos. 15-38-21009 and 16-38-00952).
The theoretical study by the shortest-graph method is supported by Russian Science Foundation, Project No. 14-43-00053. 
The present position of SAK at Aix Marseille University is supported by the A*MIDEX project (Nr.~ANR-11-IDEX-0001-02) funded by the French Government ``Investissements d'Avenir'' program managed by the French National Research Agency (ANR).

\begin{widetext}
\appendix
\section{MD results}
\label{Appendix}

In the Appendix, we summarize main results from MD simulations performed in this study. Table \ref{Table1} reports the reduced excess energies and pressures at different state points in the fluid phase. Table  \ref{Table2} summarizes the values of the anharmonic correction coefficient $\beta$ evaluated using MD simulations of the crystalline phase. Finally, Tables  \ref{Table3} and  \ref{Table4} report the excess energies and pressures in the crystalline phase.

\begin{table}[h]
	\centering
	\small
	\caption{Reduced excess energy $u_{\rm ex}$ and pressure $p$ of two-dimensional Yukawa fluids evaluated using MD simulations for various coupling ($\Gamma$) and screening ($\kappa$) parameters.}
	\label{Table1}
		\begin{tabular}{l c c c c c c c c c c c c}\hline\hline
		\multicolumn{13}{c}{ $\kappa=0.5$}\\ \hline
		$\Gamma$ & 135.420 & 86.7254 & 52.7787 & 32.1811 & 19.6073 & 11.9310 & 7.27175 & 4.43126 & 2.69848 & 1.64302 & 1.00136 &  0.5  \\
		$u_{\mathrm{ex}}$ & 152.944 & 98.3115 & 60.1901 & 37.0087 & 22.8180 & 14.1176 & 8.79838 & 5.51964 & 3.48587 & 2.21772 & 1.42021 & 0.76495\\
		$p$ & 199.434 & 128.303 & 78.6946 & 48.5651 & 30.1485 & 18.8835 & 12.0216 & 7.81631 & 5.22964 & 3.63556 & 2.64961 & 1.85883\\\hline\hline
		\multicolumn{13}{c}{ $\kappa=0.6$}\\\hline
		$\Gamma$  & 140.131 & 89.5076 & 54.3171 & 32.9737 & 20.0017 & 12.1359 & 7.36665 & 4.47442 & 2.71053 & 1.64677 & 1.00106 & 0.5  \\
		$u_{\mathrm{ex}}$ & 116.984 & 75.1128 & 45.9415 & 28.2016 & 17.3768 & 10.7727 & 6.73045 & 4.24422 & 2.69421 & 1.72956 & 1.11776 & 0.61083\\
		$p$ 					& 160.369 & 103.050  & 63.1652  & 38.9451  & 24.1971  & 15.2284  & 9.76528  & 6.42899  & 4.37128  & 3.11015  & 2.32663  & 1.69701\\\hline\hline
		\multicolumn{13}{c}{ $\kappa=0.8$}\\\hline
		$\Gamma$ & 152.277 & 96.5736 & 58.0604 & 34.9737 & 21.0334 & 12.6675 & 7.61503 & 4.58845 & 2.75830 & 1.66410 & 0.99914  & 0.5  \\
		$u_{\mathrm{ex}}$ & 74.6424 & 47.7340  & 29.0608 & 17.8181 & 10.9844 & 6.84185 & 4.30139 & 2.74217 & 1.76665 & 1.15293  & 0.75437 & 0.42469\\
		$p$ 					& 112.709 & 72.1411  & 44.0441 & 27.1658 & 16.9406 & 10.7731 & 7.01845 & 4.73986 & 3.33679 & 2.47393  & 1.92983 & 1.49910\\\hline\hline		
		\multicolumn{13}{c}{ $\kappa=1.0$}\\\hline
		$\Gamma$ & 169.071 & 105.975 & 63.1038 & 37.6027 & 22.4047 & 13.3361 & 7.94729 & 4.73129 & 2.81940 & 1.68034  & 0.99956 & 0.5  \\
		$u_{\mathrm{ex}}$ & 51.5786 & 32.7335 & 19.8556 & 12.1451 & 7.50279 & 4.68984 & 2.97702 & 1.91799 & 1.25426 & 0.82932 & 0.55059 & 0.31770\\
		$p$ 	& 85.4036 & 54.2492 & 33.0215 & 20.3527 & 12.7618 & 8.19406 & 5.44279 & 3.76791 & 2.74103 & 2.10336 & 1.70075 & 1.38135\\\hline\hline		
		\multicolumn{13}{c}{ $\kappa=1.2$}\\\hline
		$\Gamma$ & 191.126 & 118.398 & 69.6429 & 40.9597 & 24.1083 & 14.1893 & 8.34919 & 4.90490 & 2.88868 & 1.70019 & 0.99984 & 0.5  \\
		$u_{\mathrm{ex}}$ & 37.5852 & 23.6918 & 14.3026 & 8.72609 & 5.39936 & 3.39637 & 2.17547 & 1.41736 & 0.93933 & 0.62908 & 0.42281 & 0.24960\\
		$p$ 	& 67.9344 & 42.8619 & 25.9838 & 16.0024 & 10.0874 & 6.56025 & 4.44041 & 3.15023 & 2.36021 & 1.86635 & 1.55301 & 1.30594\\\hline\hline		
		\multicolumn{13}{c}{ $\kappa=1.4$}\\\hline
		$\Gamma$ & 220.172 & 134.441 & 77.9949 & 45.2452 & 26.2578 & 15.2219 & 8.83634 & 5.12702 & 2.97137 & 1.72440 & 1.00140  & 0.5  \\
		$u_{\mathrm{ex}}$ & 28.5555 & 17.8503 & 10.7244 & 6.53392 & 4.05300 & 2.56405 & 1.65932 & 1.09552 & 0.73364 & 0.49726 & 0.33718 & 0.20253\\
		$p$ 	& 56.0915 & 35.0963 & 21.1892 & 13.0574 & 8.28303 & 5.45288 & 3.76392 & 2.73780 & 2.10241 & 1.70540 & 1.45171 & 1.25396\\\hline\hline		
		\multicolumn{13}{c}{ $\kappa=1.6$}\\\hline
		$\Gamma$ & 258.433 & 155.296 & 88.6297 & 50.6106 & 28.9099 & 16.4928 & 9.41249 & 5.37870 & 3.07317 & 1.75217 & 0.99889  & 0.5  \\
		$u_{\mathrm{ex}}$ & 22.4535 & 13.9136 & 8.31218 & 5.05719 & 3.14728 & 2.00498 & 1.30903 & 0.87473 & 0.59391 & 0.40446 & 0.27520 & 0.16486\\
		$p$ & 47.7294 & 29.6021 & 17.7849 & 10.9674 & 7.00739 & 4.67522 & 3.28559 & 2.44432 & 1.92230 & 1.58965 & 1.37647 & 1.15781\\\hline\hline	
		\multicolumn{13}{c}{ $\kappa=1.8$}\\\hline
		$\Gamma$ & 308.935 & 182.395 & 102.261 & 57.3435 & 32.1483 & 18.0355 & 10.1029 & 5.67241 & 3.17978 & 1.78359 & 0.99997  & 0.5  \\
		$u_{\mathrm{ex}}$ & 18.1745 & 11.1626 & 6.63304 & 4.02868 & 2.51560 & 1.61389 & 1.06328 & 0.71747 & 0.49051 & 0.33739 & 0.23058 & 0.14359\\
		$p$ 	& 41.6428 & 25.5932 & 15.3055 & 9.44338 & 6.07675 & 4.10949 & 2.93845 & 2.22906 & 1.78546 & 1.50402 & 1.32125 & 1.18748\\\hline\hline	
		\multicolumn{13}{c}{ $\kappa=2.0$}\\\hline
		$\Gamma$ & 375.818 & 217.422 & 119.600 & 65.7745 & 36.1611 & 19.8980 & 10.9232 & 6.01199 & 3.30681 & 1.81767 & 1.00051 & 0.5  \\
		$u_{\mathrm{ex}}$ & 15.0964 & 9.17319 & 5.42177 & 3.29200 & 2.06139 & 1.33276 & 0.88426 & 0.60261 & 0.41513 & 0.28650 & 0.19651 & 0.12379\\
		$p$ 					& 37.1333 & 22.5775 & 13.4413 & 8.30684 & 5.38337 & 3.68921 & 2.67835 & 2.06727 & 1.68347 & 1.43752 & 1.27850 & 1.16494\\\hline\hline	
		\multicolumn{13}{c}{ $\kappa=2.2$}\\\hline
		$\Gamma$ & 463.975 & 262.948 & 141.568 & 76.2338 & 41.0173 & 22.0958 & 11.9035 & 6.41082 & 3.45056 & 1.85303 & 1.00113 & 0.5  \\
		$u_{\mathrm{ex}}$ & 12.7875 & 7.69994 & 4.52708 & 2.74830 & 1.72461 & 1.12217 & 0.75368 & 0.51642 & 0.35777 & 0.24734 & 0.17009 & 0.10850\\
		$p$ 					& 33.6575 & 20.2710 & 12.0118 & 7.43585 & 4.85060 & 3.36425 & 2.48426 & 1.94445 & 1.60450 & 1.38520 & 1.24473 & 1.14408\\\hline\hline
		\multicolumn{13}{c}{ $\kappa=2.4$}\\\hline
		$\Gamma$ & 578.968 & 320.871 & 168.949 & 89.0382 & 46.8778 & 24.7092 & 12.9953 & 6.85634 & 3.60307 & 1.89919 & 0.99952 & 0.5  \\
		$u_{\mathrm{ex}}$ & 10.9709 & 6.56430 & 3.83850 & 2.33031 & 1.47100 & 0.96365 & 0.65089 & 0.44974 & 0.31141 & 0.21697 & 0.14862 & 0.09589\\
		$p$ & 30.8215 & 18.4175 & 10.8648 & 6.74135 & 4.43655 & 3.11369 & 2.32748 & 1.84722 & 1.53931 & 1.34446 & 1.21673 & 1.12942\\\hline\hline		
		\multicolumn{13}{c}{ $\kappa=2.6$}\\\hline
		$\Gamma$ & 723.656 & 392.384 & 202.051 & 104.080 & 53.5742 & 27.6270 & 14.2191 & 7.32182 & 3.76653 & 1.93971 & 1.00200 & 0.5  \\
		$u_{\mathrm{ex}}$ & 9.50055 & 5.63818 & 3.28596 & 1.99866 & 1.26783 & 0.83500 & 0.56905 & 0.39442 & 0.27600 & 0.19145 & 0.13130 & 0.08576\\
		$p$ 	& 28.3633 & 16.8096 & 9.89231 & 6.16190 & 4.09049 & 2.90245 & 2.19936 & 1.76426 & 1.48858 & 1.30961 & 1.19408 & 1.11954\\\hline\hline
		\multicolumn{13}{c}{ $\kappa=2.8$}\\\hline
		$\Gamma$ & 893.746 & 474.549 & 239.143 & 120.685 & 60.8483 & 30.6642 & 15.4796 & 7.80951 & 3.93161 & 1.98042 & 1.00296 & 0.5  \\
		$u_{\mathrm{ex}}$ & 8.19448 & 4.82859 & 2.81518 & 1.71951 & 1.09985 & 0.73051 & 0.50093 & 0.35038 & 0.24489 & 0.17117 & 0.11700 & 0.07671\\
		$p$ 	& 25.9004 & 15.2521 & 8.98792 & 5.63831 & 3.78782 & 2.72194 & 2.08856 & 1.69631 & 1.44344 & 1.28133 & 1.17497 & 1.10201\\\hline\hline
		\multicolumn{13}{c}{ $\kappa=3.0$}\\\hline
		$\Gamma$ & 1071.02 & 558.495 & 276.444 & 136.953 & 67.7922 & 33.5897 & 16.6383 & 8.22716 & 4.07874 & 2.02013 & 0.99949 & 0.5  \\
		$u_{\mathrm{ex}}$ & 6.93189 & 4.07091 & 2.38838 & 1.47193 & 0.95056 & 0.64023 & 0.44340 & 0.31146 & 0.21994 & 0.15395 & 0.10494 & 0.06958\\
		$p$ 	& 23.1181 & 13.5906 & 8.07317 & 5.12679 & 3.49444 & 2.55590 & 1.98879 & 1.63334 & 1.40554 & 1.25677 & 1.15868 & 1.09682\\\hline\hline
		\end{tabular}
\end{table}

\begin{table}[!t]
	\centering
	\small
	\caption{Values of the anharmonic correction coefficient $\beta$ for different screening parameter $\kappa$.}
	\label{Table2}
		\begin{tabular}{l c c c c c c c c c c c c c c c c c}
		$\kappa$ & 0.0 & 0.2 & 0.3 & 0.4 & 0.6 & 0.8 & 1.0 & 1.2 & 1.4 & 1.6 & 1.8 & 2.0 & 2.2 & 2.4 & 2.6 & 2.8 & 3.0 \\\hline
		$\beta(\kappa)$	& 3.01 & 9.23 & 12.38 & 14.30 & 10.53 & 9.71 & 9.35 & 9.28 & 9.14 & 9.08 & 8.97 & 8.855 & 8.68 & 8.71 & 8.46 & 8.47 & 8.51
		\end{tabular}
\end{table}

\begin{table}[!t]
	\centering
	\small
	\caption{Reduced excess energy $u_{\mathrm{ex}}$ of the 2D Yukawa crystal obtained in MD simulations for various screening parameters $\kappa$ and reduced coupling parameters $\Gamma_{\rm m}/\Gamma$.}
	\label{Table3}
		\begin{tabular}{lccccccccc}
		\multicolumn{1}{c|}{ $\kappa$}& \multicolumn{9}{c}{$\Gamma_{\rm m}/\Gamma$}     \\ \hline\hline
		\multicolumn{1}{l|}{ }& 0.1 & 0.2 & 0.3 & 0.4 & 0.5 & 0.6 & 0.7 & 0.8 & 0.9 \\ \cline{2-10}
		\multicolumn{1}{l|}{0.5} & 1595.62 & 798.828 & 532.689 & 399.681 & 319.981 & 266.796 & 228.880 & 200.332 & 178.283 \\
		\multicolumn{1}{l|}{0.6} & 1217.36 & 609.282 & 406.628 & 305.117 & 244.469 & 203.938 & 174.914 & 153.188 & 136.267 \\
		\multicolumn{1}{l|}{0.8} & 773.025 & 387.104 & 258.328 & 194.074 & 155.484 & 129.733 & 111.343 & 97.5607 & 86.8364 \\
		\multicolumn{1}{l|}{1.0} & 529.643 & 265.306 & 177.235 & 133.215 & 106.726 & 89.1490 & 76.5169 & 67.1314 & 59.7831 \\
		\multicolumn{1}{l|}{1.2} & 382.522 & 191.740 & 128.152 & 96.3972 & 77.2970 & 64.6022 & 55.5318 & 48.7317 & 43.4438   \\
		\multicolumn{1}{l|}{1.4} & 287.408 & 144.232 & 96.4804 & 72.5942 & 58.2862 & 48.7586 & 41.9386 & 36.8484 & 32.8838   \\
		\multicolumn{1}{l|}{1.6} & 223.185 & 112.096 & 75.0671 & 56.5515 & 45.4466 & 38.0606 & 32.7681 & 28.8120 & 25.7391   \\
		\multicolumn{1}{l|}{1.8} & 178.133 & 89.6228 & 60.0889 & 45.3116 & 36.4631 & 30.5563 & 26.3521 & 23.1896 & 20.7451   \\
		\multicolumn{1}{l|}{2.0} & 145.774 & 73.3800 & 49.2712 & 37.2003 & 29.9641 & 25.1447 & 21.7011 & 19.1314 & 17.1275   \\
		\multicolumn{1}{l|}{2.2} & 121.609 & 61.3067 & 41.2021 & 31.1620 & 25.1352 & 21.1177 & 18.2517 & 16.1113 & 14.4385   \\
		\multicolumn{1}{l|}{2.4} & 102.908 & 51.9465 & 34.9672 & 26.4819 & 21.3920 & 17.9999 & 15.5706 & 13.7650 & 12.3602   \\
		\multicolumn{1}{l|}{2.6} & 87.4157 & 44.2324 & 29.8212 & 22.6181 & 18.2990 & 15.4212 & 13.3710 & 11.8300 & 10.6351   \\
		\multicolumn{1}{l|}{2.8} & 73.5771 & 37.3025 & 25.2028 & 19.1490 & 15.5271 & 13.1108 & 11.3865 & 10.0997 & 9.10597   \\
		\multicolumn{1}{l|}{3.0} & 60.2002 & 30.6118 & 20.7457 & 15.8118 & 12.8497 & 10.8840 & 9.47465 & 8.43053 & 7.65187   \\
		\end{tabular}
\end{table}

\begin{table}[!t]
	\centering
	\small
	\caption{Reduced pressure (compressibility)  $p$ of the 2D Yukawa crystal obtained in MD simulations for various screening parameters $\kappa$ and reduced coupling parameters $\Gamma_{\rm m}/\Gamma$.}
	\label{Table4}
		\begin{tabular}{lccccccccc}
		\multicolumn{1}{c|}{ $\kappa$}& \multicolumn{9}{c}{$\Gamma_{\rm m}/\Gamma$}     \\ \hline\hline
		\multicolumn{1}{l|}{ }& 0.1 & 0.2 & 0.3 & 0.4 & 0.5 & 0.6 & 0.7 & 0.8 & 0.9 \\ \cline{2-10}
		\multicolumn{1}{l|}{0.5} & 2080.63 & 1041.70 & 694.789 & 521.370 & 417.454 & 348.100 & 298.669 & 261.442 & 232.679 \\
		\multicolumn{1}{l|}{0.6} & 1669.06 & 835.485 & 557.680 & 418.523 & 335.380 & 279.814 & 240.022 & 210.233 & 187.022 \\
		\multicolumn{1}{l|}{0.8} & 1168.03 & 585.024 & 390.480 & 293.406 & 235.104 & 196.197 & 168.410 & 147.583 & 131.370 \\
		\multicolumn{1}{l|}{1.0} & 878.208 & 440.005 & 294.000 & 221.023 & 177.106 & 147.964 & 127.016 & 111.450 & 99.2542 \\
		\multicolumn{1}{l|}{1.2} & 693.046 & 347.470 & 232.288 & 174.765 & 140.162 & 117.162 & 100.726 & 88.4011 & 78.8053   \\
		\multicolumn{1}{l|}{1.4} & 566.555 & 284.386 & 190.275 & 143.196 & 114.994 & 96.2113 & 82.7636 & 72.7234 & 64.8975   \\
		\multicolumn{1}{l|}{1.6} & 476.692 & 239.477 & 160.406 & 120.865 & 97.1465 & 81.3696 & 70.0608 & 61.6053 & 55.0288   \\
		\multicolumn{1}{l|}{1.8} & 410.580 & 206.621 & 138.561 & 104.505 & 84.1086 & 70.4915 & 60.7970 & 53.5005 & 47.8555   \\
		\multicolumn{1}{l|}{2.0} & 361.191 & 181.859 & 122.134 & 92.2267 & 74.2973 & 62.3524 & 53.8144 & 47.4405 & 42.4641   \\
		\multicolumn{1}{l|}{2.2} & 322.729 & 162.732 & 109.386 & 82.7430 & 66.7485 & 56.0825 & 48.4703 & 42.7821 & 38.3327   \\
		\multicolumn{1}{l|}{2.4} & 291.498 & 147.173 & 99.0847 & 75.0489 & 60.6307 & 51.0175 & 44.1300 & 39.0087 & 35.0158   \\
		\multicolumn{1}{l|}{2.6} & 263.437 & 133.325 & 89.9002 & 68.1935 & 55.1747 & 46.4976 & 40.3128 & 35.6615 & 32.0486   \\
		\multicolumn{1}{l|}{2.8} & 235.188 & 119.260 & 80.5872 & 61.2342 & 49.6540 & 41.9257 & 36.4074 & 32.2829 & 29.0897   \\
		\multicolumn{1}{l|}{3.0} & 203.533 & 103.516 & 70.1601 & 53.4777 & 43.4588 & 36.8063 & 32.0351 & 28.4887 & 25.8081  \\
		\end{tabular}
\end{table}
\end{widetext}

\bibliography{Ref-2D-Yukawa} 

\end{document}